\documentclass[prl,letterpaper,nobalancelastpage,twocolumn,nofootinbib]{revtex4}

\usepackage{graphicx}
\usepackage{amsmath}
\usepackage{amssymb}
\usepackage[english]{babel}
\usepackage{color}
\usepackage[version=4]{mhchem}
\usepackage[hidelinks]{hyperref}
\usepackage{dsfont}

\usepackage{soul}
\newcommand{\ket}[1]{| #1 \rangle}

\newcommand{\Caltech}{Division of Physics, Mathematics and Astronomy, California Institute of Technology, Pasadena, CA 91125, USA}

\renewcommand{\cite}[1]{\mbox{\citep{#1}}}

\begin{document}
\title{2000-times repeated imaging of strontium atoms in clock-magic tweezer arrays}
\author{Jacob P. Covey}
\affiliation{\Caltech}
\author{Ivaylo S. Madjarov}
\affiliation{\Caltech}
\author{Alexandre Cooper}
\affiliation{\Caltech}
\author{Manuel Endres}\email{mendres@caltech.edu}
\affiliation{\Caltech}

\begin{abstract}
We demonstrate single-atom resolved imaging with a survival probability of $0.99932(8)$ and a fidelity of $0.99991(1)$, enabling us to perform repeated high-fidelity imaging of single atoms in tweezers for thousands of times. We further observe lifetimes under laser cooling of more than seven minutes, an order of magnitude longer than in previous tweezer studies. Experiments are performed with strontium atoms in $813.4~\text{nm}$ tweezer arrays, which is at a magic wavelength for the clock transition. Tuning to this wavelength is enabled by off-magic Sisyphus cooling on the intercombination line, which lets us choose the tweezer wavelength almost arbitrarily. We find that a single not retro-reflected cooling beam in the radial direction is sufficient for mitigating recoil heating during imaging. Moreover, this cooling technique yields temperatures below $5~\mu$K, as measured by release and recapture. Finally, we demonstrate clock-state resolved detection with average survival probability of $0.996(1)$ and average state detection fidelity of $0.981(1)$. Our work paves the way for atom-by-atom assembly of large defect-free arrays of alkaline-earth atoms, in which repeated interrogation of the clock transition is an imminent possibility.
\end{abstract}
\maketitle

Optical lattice clocks of alkaline-earth(-like) atoms (AEAs) have reached record precision~\cite{Ludlow2015,Campbell2017} for which the exploration of fundamental physics, such as geodesy~\cite{Grotti2018}, gravitational waves~\cite{Kolkowitz2016}, and even dark matter~\cite{Wcislo2016} is now a possibility. Yet, despite the precise optical control of AEAs that has been demonstrated in a low-entropy array~\cite{Campbell2017}, the ability to address and detect single atoms is currently lacking. Such single-atom control techniques would provide new avenues for optical clock systems. Specifically, they are required for realizing a myriad of quantum computing protocols for AEAs using clock states~\cite{Daley2008,Gorshkov2009,Daley2011,Daley2011e,Pagano2018} and could provide the foundation for generating and probing entanglement for quantum-enhanced metrology~\cite{Gil2014,Norcia2018a}. Optical tweezer (OT) techniques have matured into a powerful tool for single-atom control, e.g., they provide the versatility required for atom-by-atom assembly of defect-free arrays~\cite{Barredo2016,Endres2016,Kim2018, Kumar2018, Robens2017} and they automatically position single atoms at distances such that interaction shifts on the clock-transition are expected to be strongly reduced~\cite{Chang2004}. Further, OT experiments generally have fast experimental repetition rates and, as demonstrated below, enable repeated low-loss clock-state read-out without reloading atoms. Such techniques could provide a pathway for quasi-continuous interleaved clock operation in order to tame the Dick effect~\cite{Schioppo2017}. Since state-insensitive `magic' trapping conditions are required for clock operation~\cite{Ye2008}, tweezers operating at a clock-magic wavelength are highly suited for these directions. 

\begin{figure}[t!]
	\centering
	\includegraphics[width=\columnwidth]{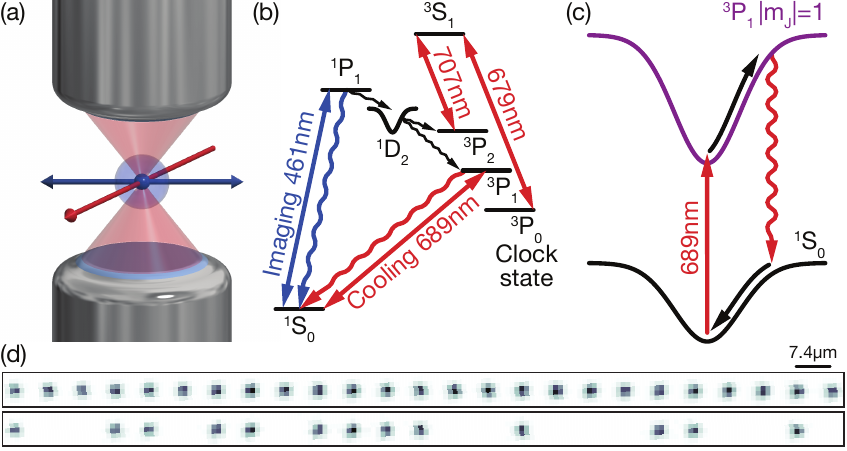}
	\caption{
	\textbf{Low-loss imaging scheme for $^{88}$Sr.}
	{(a)}~Our single atom imaging scheme requires only a single \textit{not} retro-reflected cooling beam at $689$ nm for narrow-line Sisyphus cooling (red single arrow) to compensate recoil heating from fluorescence generated by exciting atoms with a single retro-reflected $461$ nm beam (blue double arrow). Microscope objectives with $\text{NA}=0.5$ are used to generate $813.4$ nm tweezers and to collect the  fluorescence light. {(b)} Previous studies found an imaging loss channel via the decay of $\ce{^{1}P_{1}}$ to $\ce{^{1}D_{2}}$ with a branching ratio of $\sim 1:20000$, where atoms left the tweezers since $\ce{^{1}D_{2}}$ was strongly anti-trapped~\cite{Cooper2018a,Norcia2018b}. Crucially, the $\ce{^{1}D_{2}}$ state is now trapped in $813.4$ nm and our results show that atoms are recovered into the $\ce{^{3}P_{J}}$ manifold with very high probability. Two lasers ($679$ nm and $707$ nm) repump atoms to $\ce{^{3}P_{1}}$, which decays back into the ground state, thus closing the $\ce{^{1}D_{2}}$ loss channel. {(c)}~We use narrow-line \textit{attractive} Sisyphus cooling on the $m_J=\pm1$ states~\cite{Chen2018}, originally proposed in Refs.~\cite{Taieb1994,Ivanov2011}. This mechanism is based on the excited state being more strongly trapped than the ground state (in contrast to \textit{repulsive} Sisyphus cooling demonstrated recently by us~\cite{Cooper2018a}). Atoms at the bottom of the trap are excited and have to climb up a steeper potential than they would in the ground state, leading to an average reduction in energy after spontaneous emission. Cooling results from the trapping potential mismatch, and not from photon recoil, thus requiring only a single cooling beam. {(d)}~Average image (top) and single-shot image (bottom) of atomic fluorescence from twenty-five uniformized tweezers with an imaging time of one second.
	}
	\label{FigOverview}

\end{figure}

Recently, two-dimensional arrays of AEAs, specifically Sr~\cite{Cooper2018a,Norcia2018b} and Yb~\cite{Saskin2018}, in optical tweezers have been demonstrated, including single-atom resolved imaging. Cooling during imaging has been performed on the narrow $\ce{^{1}S_{0}}\leftrightarrow\ce{^{3}P_{1}}$ intercombination line (see Fig.~\ref{FigOverview}), similar to quantum gas microscopes for Yb~\cite{Yamamoto2016}. To this end, trapping wavelengths have been chosen such that the differential polarizability on this transition is small, enabling motional sidebands to be spectrally resolved in the case of Sr~\cite{Cooper2018a,Norcia2018b}, but precluding the possibility of achieving a magic trapping condition for the optical clock transition. Significantly, a more versatile Sisyphus cooling mechanism~\cite{Wineland1992} has recently been observed for Sr atoms~\cite{Cooper2018a,Chen2018}, providing a general pathway for cooling on narrow lines with strong polarizability mismatch. This observation combined with prior predictions~\cite{Taieb1994, Ivanov2011} should allow for tweezer trapping and cooling of AEAs -- and more generally atoms with narrow transitions -- in a very wide range of wavelengths. 

Here, we demonstrate detection and cooling of single $^{88}$Sr atoms in clock-magic optical tweezer arrays of wavelength $813.4$ nm~\cite{Katori2003,Takamoto2003,Takamoto2005,Boyd2006,Campbell2017,Norcia2018a} where the loss during imaging is suppressed by two orders of magnitude compared to previous work for Sr~\cite{Cooper2018a,Norcia2018b}. Specifically, we find a survival probability of $0.99932(8)$ and a fidelity of $0.99991(1)$ for single atom detection, enabling us to perform repeated high-fidelity detection for thousands of times. We also observe lifetimes under laser cooling of more than seven minutes.

These values provide a benchmark for simultaneous low-loss  and high-fidelity imaging as well as trapping lifetimes for single neutral atoms, including work with alkalis~\cite{Kuhr2016, Endres2016, Barredo2016, Kim2018, Kumar2018, Robens2017, Wu2018,Kwon2017}. We expect this development to be important for improved scalability of atom-by-atom assembly schemes~\cite{Kuhr2016,Endres2016,Barredo2016, Kim2018, Kumar2018, Robens2017} and for verifying high-fidelity quantum operations with neutral atoms~\cite{Levine2018, Wu2018}. For example, the success probability in atom-by-atom assembly is fundamentally limited by $p_s^M$, where $p_s$ is the combined survival probability for two images and hold time for rearrangement, and $M$ is the final array size~\cite{Endres2016}. Our work improves this fundamental limitation of $p_s^M\sim 0.5$ to $M \gtrsim 1000$, enabling in principle assembly of arrays with thousands of atoms in terms of imaging- and vacuum-limited lifetimes. Finally, we demonstrate single-shot clock-state resolved detection with average fidelity of $0.981(1)$ and average atom survival probability of $0.996(1)$, which could be used for repeated clock interrogation and readout periods without reloading.\\  

\textit{Experimental techniques -} Single atoms are loaded stochastically from a narrow-line magneto-optical trap into an array of tweezers as described in detail in Ref.~\cite{Cooper2018a}. In contrast to our previous work, we use 813.4 nm light to generate tweezers (Fig.~\ref{FigOverview}a). While providing a magic-wavelength for the clock transition, this wavelength also closes a previously observed loss channel, providing the basis for the low-loss detection demonstrated here (Fig.~\ref{FigOverview}b)~\cite{Cooper2018a,Norcia2018b}. Further, the imaging scheme is simplified to a single \textit{not} retro-reflected cooling beam at 689 nm and a retro-reflected imaging beam at 461 nm. Both beams propagate in the plane orthogonal to the tweezer propagation axis. The cooling (imaging) beam is polarized parallel (perpendicular) to the tweezer propagation axis. We modulate the retro-mirror of the imaging beam to wash out interference patterns~\cite{Endres2016}. The tweezers are linearly polarized and have a depth of $\approx450$~$\mu$K and waist of $\approx700$~nm. The array of 25 tweezers has a  spacing of $\approx7.4$~$\mu$m and is uniformized to within $\approx2\%$~\cite{Cooper2018a}. Tweezer arrays are generated with a bottom objective, while a second top objective is used to image both the tweezer light on a diagnostic camera and the fluorescence light on an electron multiplied charge-coupled device (EMCCD) camera. 

\begin{figure}[t!]
	\centering
	\includegraphics[width=\columnwidth]{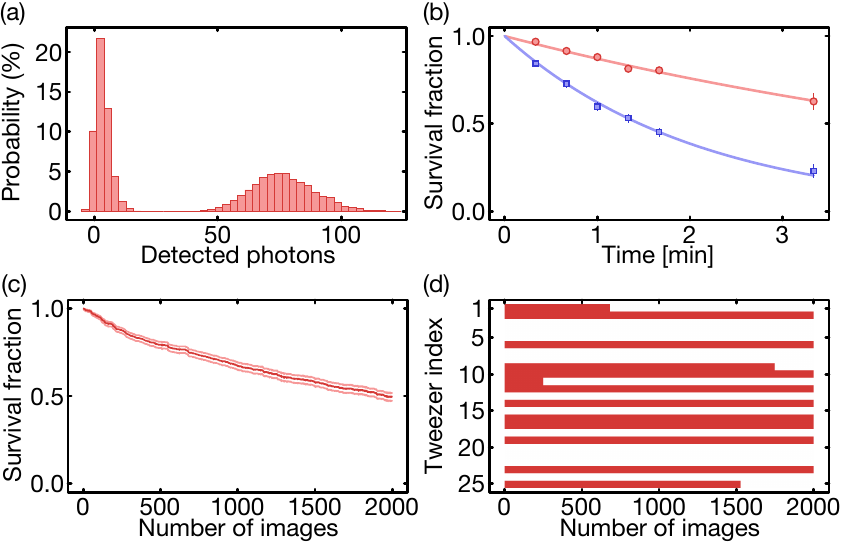}
	\caption{
	\textbf{Low-loss high-fidelity imaging.}
	{(a)}~Histogram of fluorescence counts from a single representative tweezer. We find a detection fidelity of 0.99991(1) and an average survival probability of 0.99932(8), demonstrating simultaneous high-fidelity and low-loss imaging. Results are for an imaging time of $t=50$ ms under simultaneous repumping and Sisyphus cooling. 
	{(b)}~The survival fraction as a function of hold time in minutes under these imaging conditions (blue squares and fitted line). Importantly, we find a lifetime of $\tau=126(3)$~s, while only needing $t \lesssim 50$~ms imaging time for reaching high detection fidelity, leading to small loss fractions consistent with $\exp(-t/\tau)$. Moreover, we find a lifetime of $434(13)$ s under Sisyphus cooling alone (without $461$ nm) demonstrating a  vacuum-limited lifetime greater than seven minutes (red circles and fitted line).
	{(c)}~Survival fraction versus image number for 2000 repeated images. The dark red line represents the mean over 40 realizations, with the lighter red lines showing the standard error of the mean. Atoms are imaged with high fidelity for 50~ms followed by a 29~ms cooling block. 
	{(d)}~A representative realization of atom detection over the course of the 2000 images. Detected atoms are plotted in red versus the image number, where the rows represent the 25 tweezers. Note that we find no occurrences of atoms returning after they are lost. A video of the 2000 images is available online.
	}
	\label{FigImagingLifetime}
\end{figure}

Cooling during imaging is based on a narrow-line \textit{attractive} Sisyphus cooling scheme on the $7.4$~kHz transition at $689$ nm (Fig.~\ref{FigOverview}c), following the original proposals in Refs.~\cite{Taieb1994, Ivanov2011}, which has been observed  only very recently in continuous beam deceleration~\cite{Chen2018}. In contrast to the \textit{repulsive} Sisyphus cooling scheme demonstrated recently by our group~\cite{Cooper2018a}, the attractive scheme relies on the excited state experiencing a significantly stronger trapping potential than the ground state. For linearly polarized trapping light, this situation is realized for the the $m_J=\pm1$ sublevels of $\ce{^{3}P_{1}}$ in wavelengths ranging from $\approx700~\text{nm}$ to $\approx900~\text{nm}$. (For longer wavelengths, including $1064$ nm, repulsive Sisyphus cooling can be used.) This enables us to fine-tune the wavelength to $813.4$ nm, which is magic for the clock transition to $\ce{^{3}P_{0}}$, while providing cooling conditions for the transition to $\ce{^{3}P_{1}}$.\\ 

\textit{Imaging results -} Our results show simultaneous high-fidelity and low-loss detection of single atoms. First, we observe a histogram of photons collected in $50$~ms with clearly resolved count distributions corresponding to cases with no atom and a single atom (Fig.~\ref{FigImagingLifetime}a). Taking a second image, after $29$~ms hold time under Sisyphus cooling alone, we find a survival probability of 0.99932(8). At the same time, the fidelity of the scheme (defined by the accuracy of distinguishing no atom from a single atom~\cite{Cooper2018a}) reaches a value of 0.99991(1), demonstrating simultaneous low-loss and high-fidelity imaging. These values are enabled by a lifetime under imaging conditions of more than two minutes (Fig.~\ref{FigImagingLifetime}b), while requiring only tens of milliseconds to acquire enough photons. We also find lifetimes under Sisyphus cooling (without $461$~nm light) that are longer than seven minutes.

These results enable us to repeatably image single atoms for thousands of times. Specifically, we alternate between $50$~ms long imaging blocks and $29$~ms pure cooling blocks for 2000 times, and collect photons on the \mbox{EMCCD} camera in each imaging block. Recording the survival probability as a function of the number of images $N$, we find that even after 2000 high-fidelity images, the survival fraction stays above $\approx 0.5$. In more detail, the decay follows an approximate exponential trend with $p_1^N$, where the single image survival probability is $p_1 \approx 0.9997$, slightly higher than the above quoted value measured with only two images. We note that the success probability in atom-by-atom assembly schemes is also limited by $p_1^{2N}$ as a function of the final array size. (The factor of two appears because two consecutive images with interleaved hold times are needed for successful rearrangement.) Our results indicate that assembly of systems with thousands of atoms could be possible in terms of imaging fidelity, loss probability, and lifetime during the assembly step. \\

\begin{figure}[t!]
	\centering
	\includegraphics[width=\columnwidth]{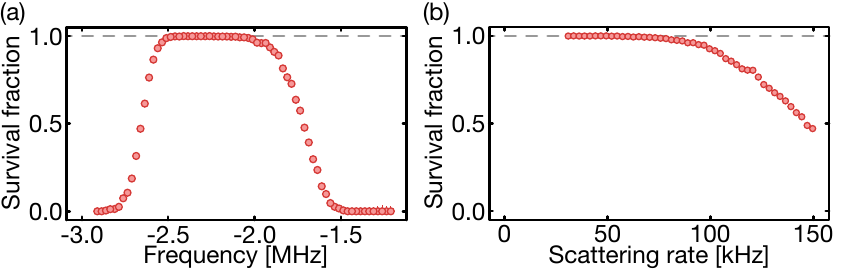}
	\caption{
	\textbf{Sisyphus cooling during imaging.} {(a)}~The survival probability of the atom versus the detuning (with respect to the free space resonance) of the $689$ nm Sisyphus cooling beam. A broad feature with high survival fraction is observed with a red-detuned edge approximately at the detuning for exciting atoms at the trap bottom, consistent with our interpretation of attractive Sisyphus cooling. Data is for a 50 ms imaging time with a $461$~nm scattering rate of $\approx 41$ kHz. For comparison, the results in Fig.~\ref{FigImagingLifetime} are for $-2.3$ MHz detuning. We perform cooling during imaging with an intensity of the $689$ nm beam of $I/I_s\approx1000$, where $I_s$ is the saturation intensity. {(b)}~The survival fraction of single atoms versus the scattering rate from the $461$ nm imaging beam under simultaneous repumping and Sisyphus cooling for an imaging time of $50~\text{ms}$. For scattering rates $\gtrsim 80$ kHz, increased loss indicates that Sisyphus cooling is not able to compensate for $461$ nm recoil heating anymore. For comparison, the results in Fig.~\ref{FigImagingLifetime} are for $\approx 41$ kHz scattering rate.}
	\label{FigCooling}
\end{figure}

\textit{Cooling results -} These low-loss high-fidelity results are achieved by optimizing the Sisyphus cooling frequency and picking a conservative $461$~nm scattering rate as detailed in Fig.~\ref{FigCooling} and the corresponding caption. 

\begin{figure}[t!]
	\centering
	\includegraphics[width=\columnwidth]{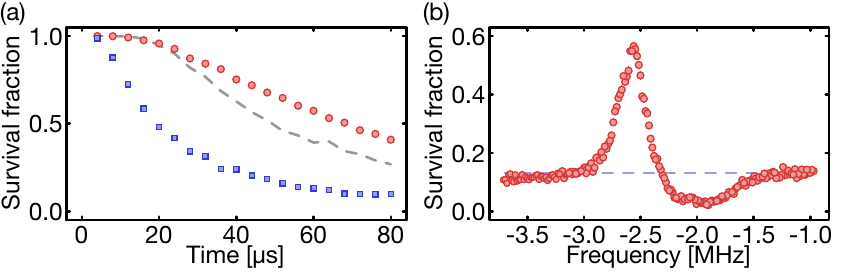}
	\caption{
	\textbf{Sisyphus cooling to low temperatures}
	{(a)}~The survival fraction in an array in a release-and-recapture measurement performed by diabatically turning off the traps for a variable time followed by a sudden switch-on~\cite{Tuchendler2008}. Lower temperatures are indicated by higher survival rates at long off-times. We show data after imaging (blue squares) and after adding a dedicated cooling block with Sisyphus cooling alone (red circles). Results are compared with classical Monte-Carlo simulations for a three-dimensional thermal distribution at $5$~$\mu$K (dashed line). Note that the release-and-recapture method is mostly sensitive to the energy distribution in the radial direction. 
	{(b)}~Survival fraction in an array after release-and-recapture for 60 $\mu$s off-time versus the red frequency during Sisyphus cooling for $25$~ms with an intensity of $I/I_s\approx200$. The dashed line represents the case without a dedicated cooling block. We find that atoms are cooled for appropriately chosen red detunings, and heated for detunings further to the blue. This is consistent with an understanding of Sisyphus cooling as an \textit{attractor} in energy space~\cite{Taieb1994, Ivanov2011}. Data in (a) is at $-2.6$~MHz detuning. 
	}
	\label{FigThermometry}
\end{figure}

In addition, attractive Sisyphus cooling without the 461 nm beam results in a radial temperature below $5~\mu$K (Fig.~\ref{FigThermometry}), which we quote as a conservative upper bound based on a release-and-recapture technique~\cite{Tuchendler2008,Thompson2013}. This technique is primarily sensitive to radial temperatures, and is compared to classical Monte-Carlo simulations to extract a temperature.  However, comparison to classical simulation would overestimate actual temperatures that are close to or below the energy scale of the radial trapping frequency ($T \lesssim \frac{\hbar \omega}{2k_B}$), which for us is roughly $2.4~\mu$K.  More precise measurement of lower temperatures could be done via resolved sideband spectroscopy, which we leave for further work. An open question in this context is whether cooling to the motional ground state can be achieved in the strongly off-magic cooling configuration used here.

As Sisyphus cooling occurs due to a trapping mismatch between ground and excited state, it is expected that cooling occurs in all directions even for a single radial cooling beam. The low loss we observe during imaging already provides evidence of this mechanism, as fluorescence recoil heating must be mitigated in all directions. Determination of the axial temperature after cooling can be achieved via techniques such as adiabatic rampdown~\cite{Alt2003,Tuchendler2008} or spectroscopy of thermally-broadened light shifts~\cite{Cooper2018a}. Our preliminary results with such techniques are consistent with three-dimensional temperatures similar to our quoted radial temperatures; however, we leave a thorough investigation to future work.  We note that we have made no explicit attempt to further cool the axial direction, and that doing so is likely possible by applying a beam in that direction. Finally, we note that the clock-magic condition of our tweezers opens the door to well-resolved sideband thermometry on the clock transition, which would be required to see resolved axial sidebands that are otherwise poorly resolved on the intercombination line at our trapping frequencies.  \\

\textit{Clock-state resolved detection -} Finally, as an outlook we characterize our ability to perform low-loss state-resolved read-out of the optical clock states $\ce{^{1}S_{0}}\equiv \ket{g}$ and $\ce{^{3}P_{0}}\equiv \ket{e}$. As detailed below, our scheme relies on shelving techniques that are routinely used in ion trap experiments to realize low-loss, high-fidelity state-resolved detection~\cite{Leibfried2003,Myerson2008,Harty2014}. They are also prevalent in optical lattice clocks with alkaline-earth atoms, but have not been extended to single-atom-resolved imaging~\cite{Takamoto2005,Boyd2006}. More generally, low-loss state-resolved detection of single \textit{neutral} atoms has been realized only recently with alkali atoms~\cite{Fuhrmanek2011,Boll2016,Kwon2017,Martinez-Dorantes2017,Wu2018}. Since in this case hyperfine states are used, simultaneous cooling during state-resolved detection in a tweezer is challenging and thus deep traps are required. This will limit scalability, and the approach has so far only been demonstrated in up to five traps~\cite{Kwon2017}. Note that Stern-Gerlach detection of hyperfine spins via spatial separation in a lattice has been performed very recently~\cite{Boll2016,Wu2018}, which provides an alternative route for high-fidelity lossless state-resolved detection. 

\begin{figure}[t!]
	\centering
	\includegraphics[width=\columnwidth]{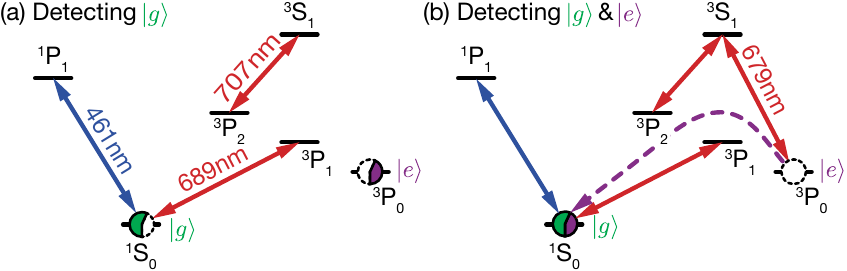}
	\caption{
	\textbf{Low-loss state-resolved detection.}
	{(a)}~A statistical mixture of optical clock qubit states is represented as a circle, where the green section at $\ce{^{1}S_{0}}\equiv|g\rangle$ represents the population in $|g\rangle$ and the purple section at $\ce{^{3}P_{0}}\equiv|e\rangle$ represents the population in $|e\rangle$. To measure the population in $|g\rangle$, we image without the 679 nm repumper during which $|e\rangle$ remains dark. The accuracy of measuring the population in $|g\rangle$ is limited both by off-resonant scattering of the tweezer light which pumps $|e\rangle$ back to $|g\rangle$, and by pathways that pump $|g\rangle$ to $|e\rangle$ such as the $\ce{^{1}D_{2}}$ decay channel and the off-resonant scattering from $\ce{^{3}P_{1}}$ during cooling. As a result, the average state detection fidelity is $0.981(1)$.
	{(b)}~We perform a second image that includes the 679 nm repumper, which pumps $|e\rangle$ to $|g\rangle$ via the $\ce{^{3}S_{1}}$ state and the 707 nm repumper, such that both states are detected. The pumping process is illustrated by the purple arrow. This image measures the population in $|g\rangle$ and $|e\rangle$, and informs whether the atom has been lost as a result of the first image. We find that the average survival probability of this double-imaging sequence is $0.996(1)$.
	}
	\label{FigStateResolved}
\end{figure}

Our scheme consists of two consecutive images (Fig.~\ref{FigStateResolved}). In the first image, we aim at detecting atoms in $\ket{g}$. To this end, we turn off the $679$ nm repump laser such that atoms in $\ket{e}$, in principle, do not scatter any photons. Hence, if we find a signal in the first image, we identify the state as $\ket{g}$. In the second image, we turn the $679$ nm repump laser back on to detect atoms in both $\ket{g}$ and $\ket{e}$. Thus, if an atom is not detected in the first image but appears in the second image, we can identify it as $\ket{e}$. If neither of the images shows a signal, we identify the state as ``no-atom''.

We find that the inaccuracy of this scheme is dominated by off-resonant scattering of the tweezer light when atoms are shelved in $|e\rangle$ during the first image. Specifically, by pumping atoms into $|e\rangle$ before imaging, we observe that, at our trap depth of $\approx450$~$\mu$K, they decay back to $|g\rangle$ with a time constant of $\tau_p=470(30)$ ms. This leads to events in the first image, where $|e\rangle$ atoms are misidentified as $|g\rangle$ atoms. To minimize the probability of misidentification, the first imaging time should be as short as possible. To reduce the imaging time, we compromise slightly on the survival probability in order to work with higher $461$~nm scattering rates in the first image (see Fig.~\ref{FigCooling}). Specifically we use $t=15$ ms at a scattering rate of $\approx72$ kHz. The second image is performed with the same settings as in Fig.~\ref{FigImagingLifetime}. 

Additionally, atoms in $|g\rangle$ can be misidentified as $|e\rangle$ if they are pumped to $|e\rangle$ in the first image. This can occur either via the $\ce{^{1}D_{2}}$ leakage channel and subsequent scattering of $707$ nm photons, or via off-resonant scattering of the trap light from $\ce{^{3}P_{1}}$ during cooling. We identify this misidentification probability by initializing atoms in $|g\rangle$, and counting how often we identify them as $|e\rangle$ in the state-resolved imaging scheme. 

In summary, we place a conservative upper-bound for the probability of misidentifying $|e\rangle$ as $|g\rangle$ by $e^{-t/\tau_p}=0.031(2)$ and we directly measure the probability of misidentifying $|g\rangle$ as $|e\rangle$ yielding $0.008(1)$. We define the average state detection infidelity for a generic initial state as the mean of these probabilities~\cite{Wu2018}, yielding an average state detection fidelity of $0.981(1)$. Further, we similarly define the average survival probability of the double imaging scheme in terms of the measured survival probabilities of $|g\rangle$ and $|e\rangle$, for which we obtain $0.996(1)$. 

Our fidelity is comparable to recent measurements with alkali atoms in tweezers~\cite{Fuhrmanek2011,Kwon2017,Martinez-Dorantes2017}, yet our survival probability is substantially higher than any tweezer- or lattice-based schemes to our knowledge~\cite{Fuhrmanek2011,Kwon2017,Martinez-Dorantes2017,Boll2016,Wu2018}. These results constitute an excellent setting for continuous measurement in an optical clock. However, we emphasize that this investigation was not exhaustive, and we expect that further optimization of these imaging parameters is possible. In general, these values could be further improved by either imaging in shallower tweezers or in tweezers at a wavelength further detuned from higher lying states. For instance, tweezers operating at 1064 nm are a promising possibility, and would be a convenient choice for operating a quantum gas microscope. Further, it is possible to switch between 813.4 nm tweezers/lattices for clock interrogation during which the trap depth can be orders of magnitude lower, and 1064 nm tweezers/lattices for imaging.

In conclusion, we have addressed two major limitations preventing the implementation of tweezer arrays for optical clock-based quantum information processing and metrology. By working at the magic wavelength for clock operation, we observe imaging with a fidelity of $0.99991(1)$ and a survival probability of $0.99932(8)$, and lifetimes under cooling of more than seven minutes. By employing a double imaging technique with specific combinations of repump lasers, we study low-loss state-resolved detection and observe an average fidelity of $0.981(1)$ with an average survival probability of $0.996(1)$. This work provides a setting for continuous measurement in an optical clock which can suppress laser fluctuations due to the Dick effect~\cite{Lodewyck2009,Schioppo2017}. Clock operation on bosonic isotopes of AEAs such as $^{88}$Sr used in this work has been performed with Sr~\cite{Akatsuka2010} and Yb~\cite{Taichenachev2006a,Barber2006}.  Moreover, the tools developed in this work enable excitation to highly-excited Rydberg states in the $\ce{^{3}S_{1}}$ series via the clock state $\ce{^{3}P_{0}}$. Engineering long-range Rydberg-mediated interactions will facilitate the generation of entanglement between optical clock qubits, which can be used for quantum information processing~\cite{Saffman2010}, quantum simulation~\cite{Labuhn2016, Bernien2017}, and quantum-enhanced metrology via spin squeezing~\cite{Gil2014}.

We acknowledge funding provided by the Institute for Quantum Information and Matter, an NSF Physics Frontiers Center (NSF Grant PHY-1733907). This work was supported by the NSF CAREER award (number 1753386), the Sloan Foundation, by the NASA/JPL President's and Director's Fund, and by Fred Blum. JPC acknowledges support from the PMA Prize postdoctoral fellowship, and AC acknowledges support from the IQIM Postdoctoral Scholar fellowship.

\bibliographystyle{h-physrev}
\bibliography{library}

\begin{thebibliography}{10}

\bibitem{Ludlow2015}
A.~D. Ludlow, M.~M. Boyd, J.~Ye, E.~Peik, and P.~O. Schmidt,
\newblock Reviews of Modern Physics {\bf 87}, 637 (2015).

\bibitem{Campbell2017}
S.~L. Campbell {\em et~al.},
\newblock Science {\bf 358}, 90 (2017).

\bibitem{Grotti2018}
J.~Grotti {\em et~al.},
\newblock Nature Physics {\bf 14}, 437 (2018).

\bibitem{Kolkowitz2016}
S.~Kolkowitz {\em et~al.},
\newblock Physical Review D {\bf 94}, 124043 (2016).

\bibitem{Wcislo2016}
P.~Wcis{\l}o {\em et~al.},
\newblock Nature Astronomy {\bf 1}, 0009 (2016).

\bibitem{Daley2008}
A.~J. Daley, M.~M. Boyd, J.~Ye, and P.~Zoller,
\newblock Physical Review Letters {\bf 101}, 170504 (2008).

\bibitem{Gorshkov2009}
A.~V. Gorshkov {\em et~al.},
\newblock Physical Review Letters {\bf 102}, 110503 (2009).

\bibitem{Daley2011}
A.~J. Daley, J.~Ye, and P.~Zoller,
\newblock The European Physical Journal D {\bf 65}, 207 (2011).

\bibitem{Daley2011e}
A.~J. Daley,
\newblock Quantum Information Processing {\bf 10}, 865 (2011).

\bibitem{Pagano2018}
G.~Pagano, F.~Scazza, and M.~Foss-Feig,
\newblock arXiv:1808.02503  (2018).

\bibitem{Gil2014}
L.~I.~R. Gil, R.~Mukherjee, E.~M. Bridge, M.~P.~A. Jones, and T.~Pohl,
\newblock Physical Review Letters {\bf 112}, 103601 (2014).

\bibitem{Norcia2018a}
M.~A. Norcia {\em et~al.},
\newblock Physical Review X {\bf 8}, 021036 (2018).

\bibitem{Barredo2016}
D.~Barredo, S.~de~Leseleuc, V.~Lienhard, T.~Lahaye, and A.~Browaeys,
\newblock Science {\bf 354}, 1021 (2016).

\bibitem{Endres2016}
M.~Endres {\em et~al.},
\newblock Science {\bf 354}, 1024 (2016).

\bibitem{Kim2018}
H.~Kim, Y.~Park, K.~Kim, H.-S. Sim, and J.~Ahn,
\newblock Physical Review Letters {\bf 120}, 180502 (2018).

\bibitem{Kumar2018}
A.~Kumar, T.-Y. Wu, F.~Giraldo, and D.~S. Weiss,
\newblock Nature {\bf 561}, 83 (2018).

\bibitem{Robens2017}
C.~Robens {\em et~al.},
\newblock Physical Review Letters {\bf 118}, 065302 (2017).

\bibitem{Chang2004}
D.~E. Chang, J.~Ye, and M.~D. Lukin,
\newblock Physical Review A {\bf 69}, 023810 (2004).

\bibitem{Schioppo2017}
M.~Schioppo {\em et~al.},
\newblock Nature Photonics {\bf 11}, 48 (2017).

\bibitem{Ye2008}
J.~Ye, H.~J. Kimble, and H.~Katori,
\newblock Science {\bf 320}, 1734 (2008).

\bibitem{Cooper2018a}
A.~Cooper {\em et~al.},
\newblock arXiv:1810.06537  (2018).

\bibitem{Norcia2018b}
M.~A. Norcia, A.~W. Young, and A.~M. Kaufman,
\newblock arXiv:1810.06626  (2018).

\bibitem{Chen2018}
C.-C. Chen, S.~Bennetts, R.~G. Escudero, F.~Schreck, and B.~Pasquiou,
\newblock arXiv:1810.07157  (2018).

\bibitem{Taieb1994}
R.~Ta{\"{i}}eb, R.~Dum, J.~I. Cirac, P.~Marte, and P.~Zoller,
\newblock Physical Review A {\bf 49}, 4876 (1994).

\bibitem{Ivanov2011}
V.~V. Ivanov and S.~Gupta,
\newblock Physical Review A {\bf 84}, 063417 (2011).

\bibitem{Saskin2018}
S.~Saskin, J.~Wilson, B.~Grinkemeyer, and J.~Thompson,
\newblock arXiv:1810.10517  (2018).

\bibitem{Yamamoto2016}
R.~Yamamoto, J.~Kobayashi, T.~Kuno, K.~Kato, and Y.~Takahashi,
\newblock New Journal of Physics {\bf 18}, 023016 (2016).

\bibitem{Wineland1992}
D.~J. Wineland, J.~Dalibard, and C.~Cohen-Tannoudji,
\newblock Journal of the Optical Society of America B {\bf 9}, 32 (1992).

\bibitem{Katori2003}
H.~Katori, M.~Takamoto, V.~G. Pal'chikov, and V.~D. Ovsiannikov,
\newblock Physical Review Letters {\bf 91}, 173005 (2003).

\bibitem{Takamoto2003}
M.~Takamoto and H.~Katori,
\newblock Physical Review Letters {\bf 91}, 223001 (2003).

\bibitem{Takamoto2005}
M.~Takamoto, F.-L. Hong, R.~Higashi, and H.~Katori,
\newblock Nature {\bf 435}, 321 (2005).

\bibitem{Boyd2006}
M.~M. Boyd {\em et~al.},
\newblock Science {\bf 314}, 1430 (2006).

\bibitem{Kuhr2016}
S.~Kuhr,
\newblock National Science Review {\bf 3}, 170 (2016).

\bibitem{Wu2018}
T.-Y. Wu, A.~Kumar, F.~G. Mejia, and D.~S. Weiss,
\newblock arXiv:1809.09197  (2018).

\bibitem{Kwon2017}
M.~Kwon, M.~F. Ebert, T.~G. Walker, and M.~Saffman,
\newblock Physical Review Letters {\bf 119}, 180504 (2017).

\bibitem{Levine2018}
H.~Levine {\em et~al.},
\newblock (2018), 1806.04682.

\bibitem{Tuchendler2008}
C.~Tuchendler, A.~M. Lance, A.~Browaeys, Y.~R.~P. Sortais, and P.~Grangier,
\newblock Physical Review A {\bf 78}, 033425 (2008).

\bibitem{Thompson2013}
J.~D. Thompson, T.~G. Tiecke, A.~S. Zibrov, V.~Vuleti{\'{c}}, and M.~D. Lukin,
\newblock Physical Review Letters {\bf 110}, 133001 (2013), 1209.3028.

\bibitem{Alt2003}
W.~Alt {\em et~al.},
\newblock Physical Review A {\bf 67}, 033403 (2003).

\bibitem{Leibfried2003}
D.~Leibfried, R.~Blatt, C.~Monroe, and D.~Wineland,
\newblock Reviews of Modern Physics {\bf 75}, 281 (2003).

\bibitem{Myerson2008}
A.~H. Myerson {\em et~al.},
\newblock Physical Review Letters {\bf 100}, 200502 (2008).

\bibitem{Harty2014}
T.~P. Harty {\em et~al.},
\newblock Physical Review Letters {\bf 113}, 220501 (2014).

\bibitem{Fuhrmanek2011}
A.~Fuhrmanek, R.~Bourgain, Y.~R.~P. Sortais, and A.~Browaeys,
\newblock Physical Review Letters {\bf 106}, 133003 (2011).

\bibitem{Boll2016}
M.~Boll {\em et~al.},
\newblock Science {\bf 353}, 1257 (2016).

\bibitem{Martinez-Dorantes2017}
M.~Martinez-Dorantes {\em et~al.},
\newblock Physical Review Letters {\bf 119}, 180503 (2017).

\bibitem{Lodewyck2009}
J.~Lodewyck, P.~G. Westergaard, and P.~Lemonde,
\newblock Physical Review A {\bf 79}, 061401 (2009).

\bibitem{Akatsuka2010}
T.~Akatsuka, M.~Takamoto, and H.~Katori,
\newblock Physical Review A {\bf 81}, 023402 (2010).

\bibitem{Taichenachev2006a}
A.~Taichenachev {\em et~al.},
\newblock Physical Review Letters {\bf 96}, 083001 (2006).

\bibitem{Barber2006}
Z.~Barber {\em et~al.},
\newblock Physical Review Letters {\bf 96}, 083002 (2006).

\bibitem{Saffman2010}
M.~Saffman, T.~G. Walker, and K.~M{\o}lmer,
\newblock Reviews of Modern Physics {\bf 82}, 2313 (2010).

\bibitem{Labuhn2016}
H.~Labuhn {\em et~al.},
\newblock Nature {\bf 534}, 667 (2016).

\bibitem{Bernien2017}
H.~Bernien {\em et~al.},
\newblock Nature {\bf 551}, 579 (2017).

\end{thebibliography}

\end{document}